\documentclass[preprint,aps,amsmath,showpacs,showkeys,nofootinbib, 
superscriptaddress]{revtex4} 
\usepackage[dvips]{graphicx} 
 
\begin{document} 

\title{\textbf{Pion-assisted charmed dibaryon candidate}} 

\author{A.~Gal} 
\email{avragal@savion.huji.ac.il} 
\affiliation{Racah Institute of Physics, The Hebrew University, 
Jerusalem 91904, Israel}

\author{H.~Garcilazo} 
\email{humberto@esfm.ipn.mx} 
\affiliation{Escuela Superior de F\' \i sica y Matem\'aticas \\ 
Instituto Polit\'ecnico Nacional, Edificio 9, 07738 M\'exico D.F., Mexico} 

\author{A.~Valcarce} 
\email{valcarce@fis.usal.es} 
\affiliation{Departamento de F\'\i sica Fundamental, \\ 
Universidad de Salamanca, E-37008 Salamanca, Spain} 

\author{T.~Fern\'andez-Caram\'es} 
\email{carames@usal.es} 
\affiliation{Departamento de F\'\i sica Fundamental, \\ 
Universidad de Salamanca, E-37008 Salamanca, Spain} 

\date{\today} 

\begin{abstract} 

The $\Lambda_c(2286)N$ system is studied in a chiral constituent quark model 
and the resulting $s$-wave interaction is used in separable form within 
three-body models of the $\pi\Lambda_c N$ system with quantum numbers 
$(C,I,J^P)=(+1,\frac{3}{2},2^+)$. Separable interactions are also used for 
the dominant $p$-wave pion-baryon channels dominated by the $\Delta(1232)$ 
and $\Sigma_c(2520)$ resonances. Faddeev equations with relativistic 
kinematics are solved on the real axis to search for bound states and 
in the complex plane to search for three-body resonances. Some of the 
models considered generate a very narrow bound state, requiring isospin 
violation for its strong decay. Other models lead to a narrow resonance 
($\Gamma \lesssim 0.4$~MeV) for resonance mass below the $\Sigma_c(2455)N$ 
threshold. This would be the lowest-lying $C=+1$ dibaryon, with mass 
estimated as $\approx$3370$\pm$15~MeV.  

\end{abstract} 




\maketitle 

\newpage 

\section{Introduction} 
\label{sec:intro} 

Several pion-assisted dibaryon candidates of the type $\pi BB'$, with 
a $p$-wave pion interacting with baryons $B$ and $B'$ that interact in 
$s$ waves, were suggested in Ref.~\cite{GAGA08}. Considered in detail 
was the $\pi\Lambda N$ system in the channel $(I,J^P)=(\frac{3}{2},2^+)$ 
which is dominated by configurations where the $p$-wave $\pi N$ 
$(I,J^P)=(\frac{3}{2},\frac{3}{2}^+)$ $\Delta(1232)$ resonance 
is coupled to an $s$-wave $\Lambda$, and the $p$-wave $\pi\Lambda$ 
$(I,J^P)=(1,\frac{3}{2}^+)$ $\Sigma(1385)$ resonance is coupled to an $s$-wave 
nucleon. The choice $(I,J^P)=(\frac{3}{2},2^+)$ ensures that the spins and 
isospins of the three hadrons are parallel, with the two baryons necessarily 
in a $^3S_1$ state, leading to maximal attraction since all spin and isospin 
recoupling coefficients in this channel are equal to one. The $(I,J^P)=
(\frac{3}{2},2^+)$ $\pi\Lambda N$--$\pi\Sigma N$ coupled-channel system 
was studied subsequently \cite{GAGA10,GAGA13}, concluding that it 
resonates some 10-20 MeV below the $\pi\Sigma N$ threshold \cite{GAGA13}. 
Other pion-assisted dibaryon candidates suggested in~\cite{GAGA08} include 
$\pi\Xi N$, $\pi\Xi\Lambda$ and $\pi\Lambda_{c} N$. In the present work we 
apply the same formalism \cite{GAGA13} to study the charmed $\pi\Lambda_c N$ 
system where one replaces the ${\frac{1}{2}}_{\rm g.s.}^+$ $\Lambda(1116)$ 
baryon by the ${\frac{1}{2}}_{\rm g.s.}^+$ $\Lambda_c(2286)$ charmed baryon, 
and the ${\frac{3}{2}}^+$ $\Sigma(1385)$ resonance by the ${\frac{3}{2}}^+$ 
$\Sigma_c(2520)$ charmed resonance. An interesting aspect of this 
$\pi\Lambda_{c} N$ system is that a bound state, if occurring, 
will decay only by isospin-violating interactions since the lowest 
isospin-conserving decay channel $\Sigma_c(2455) N$ lies $\approx$30~MeV 
above the $\pi\Lambda_c N$ threshold. 

To formulate and solve a $\pi\Lambda_c N$ three-body model one needs 
to specify the input pairwise interactions. Whereas the construction of 
$p$-wave separable interactions describing the pion-baryon $\Delta(1232)$ 
and $\Sigma_c(2520)$ resonances is straightforward, the construction of the 
necessary $s$-wave separable interaction describing the $\Lambda_c(2286)N$ 
system requires special attention. In the present exploratory study we 
neglect its coupling to the $\Sigma_c(2455)N$ system, reporting briefly on 
a straightforward application of the chiral constituent quark model (CCQM) 
within the charm sector \cite{Vij05,Car12}. This model, tuned by fitting to 
the baryon and meson spectra as well as to the $NN$ interaction, provides 
predictions for charm $C=+1$ two-hadron systems that will become testable 
in due course. For an extensive review of the CCQM, see Ref.~\cite{Val05}. 

The paper is organized as follows. The input pion-baryon phenomenological 
interactions are discussed in Sect.~\ref{sec:mB}, and the input $\Lambda_c N$ 
CCQM interactions are discussed in Sect.~\ref{sec:YN}. Results of three-body 
calculations using Faddeev equations with relativistic kinematics are given 
and discussed in Sect.~\ref{sec:res}, with conclusions drawn in the last 
Sect.~\ref{sec:concl}.

\section{Pion-baryon $p$-wave interactions} 
\label{sec:mB} 

Following the discussion of the $\pi\Lambda N$ system in Ref.~\cite{GAGA08}, 
the dominant two-body interactions in the $\pi\Lambda_c N$ system are 
the $p$-wave $\pi N$ $(I,J^P)=(\frac{3}{2},\frac{3}{2}^+)$ $\Delta(1232)$ 
and $\pi\Lambda_c$ $(I,J^P)=(1,\frac{3}{2}^+)$ $\Sigma_c(2520)$ channels, 
and the $s$-wave $\Lambda_c N$ interaction in the $I=\frac{1}{2},{^{3}S_1}$ 
channel. In this section we describe the appropriate separable-interaction 
meson-baryon models, assigning particle indices 1,2,3 to charmed-hyperons, 
nucleon and pion, respectively.

\subsection{The $\pi N$ subsystem} 
\label{sec:piN} 

The Lippmann-Schwinger equation for the pion-nucleon interaction is given 
by \cite{GAGA11}: 
\begin{eqnarray} 
t_1(p_1,p_1^\prime;\omega_0) &=& V_1(p_1,p_1^\prime)+\int_0^\infty 
{p_1^{\prime\prime}}^2 dp_1^{\prime\prime}  \nonumber \\ 
& \times & V_1(p_1,p_1^{\prime\prime})\frac{1}{\omega_0-
\sqrt{m_N^2+{p_1^{\prime\prime 2}}}-\sqrt{m_\pi^2+{p_1^{\prime\prime 2}}}+
{\rm i}\epsilon}t_1(p_1^{\prime\prime},p_1^\prime;\omega_0), 
\label{eq:A1} 
\end{eqnarray} 
so that using a separable potential 
\begin{equation} 
V_1(p_1,p_1^\prime)=\gamma_1 g_1(p_1) g_1(p_1^\prime), 
\label{eq:A2} 
\end{equation} 
one gets 
\begin{equation} 
t_1(p_1,p_1^\prime;\omega_0)= g_1(p_1) \tau_1(\omega_0) g_1(p_1^\prime), 
\label{eq:A3} 
\end{equation} 
where 
\begin{equation} 
[\tau_1(\omega_0)]^{-1} =\frac{1}{\gamma_1}-\int_0^\infty p_1^2 dp_1 
\frac{g_1^2(p_1)}{\omega_0-\sqrt{m_N^2+p_1^2}-\sqrt{m_\pi^2+p_1^2}+
{\rm i}\epsilon}. 
\label{eq:A4} 
\end{equation} 
A fit to the $P_{33}$ phase shift and scattering volume using the form factor 
\begin{equation} 
g_1(p_1)=p_1[\exp(-p_1^2/\beta_1^2)+Cp_1^2\exp(-p_1^2/\alpha_1^2)], 
\label{eq:A5} 
\end{equation} 
with parameters given in Table~\ref{tab:piN}, was shown and discussed in 
Ref.~\cite{GAGA11}. Listed in the table are also the r.m.s. radii of the 
form factors $g(p)$ in momentum space and ${\tilde g}(r)$ in coordinate space, 
where ${\tilde g}(\vec r\:)={\hat r}{\tilde g}(r)$ is the Fourier transform of 
the $p$-wave form factor $g(\vec p\:)={\hat p}g(p)$, given by 
\begin{equation} 
{\tilde g}(r) \sim \int j_1(pr)g(p)p^2dp, 
\label{eq:j_1} 
\end{equation} 
with $j_1$ the spherical Bessel function for $\ell = 1$. As elaborated 
in Ref.~\cite{GAGA11}, ${\tilde g}(r)$ is not positive definite, which 
may result in negative values of $<r^2>$. A spatial-size substitute for 
$\sqrt{<r^2>}$ is provided then by $r_0^{(\pi N)}$, the first zero of 
${\tilde g}(r)$. Both values of $\sqrt{<r^2>}$ and $r_0$ listed in 
Table~\ref{tab:piN} are seen to be close to each other, but this need not 
necessarily be the case for other subsystems, as will become evident in 
the next subsection. 
 
\begin{table}[htb] 
\caption{Fitted parameters of the $\pi N$ separable $p$-wave interaction 
(\ref{eq:A2}) with form factor $g_1(p)$ defined by Eq.~(\ref{eq:A5}). 
Listed also are values of its r.m.s. momentum $\sqrt{<p^2>_{g_1}}$ 
(in fm$^{-1}$), and r.m.s. radius $\sqrt{<r^2>_{{\tilde {g_1}}}}$ and 
zero $r_0^{(\pi N)}$ (both in fm) of the coordinate-space form factor 
$\tilde {g_1}(r)$.} 
\begin{ruledtabular} 
\begin{tabular}{ccccccc} 
$\gamma_1~({\rm fm}^4)$ & $\alpha_1~({\rm fm}^{-1})$ & 
$\beta_1~({\rm fm}^{-1})$ & $C~({\rm fm}^2)$ & 
$\sqrt{<p^2>_{g_1}}$ & $\sqrt{<r^2>_{{\tilde {g_1}}}}$ & $r_0^{(\pi N)}$ \\ 
\hline 
$-$0.075869 & 2.3668 & 1.04 & 0.23 & 4.07 & 1.47 & 1.36 \\ 
\end{tabular} 
\end{ruledtabular} 
\label{tab:piN} 
\end{table} 

The pion-nucleon amplitude in the three-body system with a $\Lambda_c$ 
as spectator is given by 
\begin{equation} 
t_1(p_1,p_1^\prime;W_0,q_1)=g_1(p_1)\tau_1(W_0,q_1) g_1(p_1^\prime), 
\label{eq:A6} 
\end{equation} 
where $W_0$ is the invariant mass of the three-body system, $q_1$ is the 
relative momentum between the spectator and the c.m. of the $\pi N$ subsystem 
and 
\begin{equation} 
[\tau_1(W_0,q_1)]^{-1} = \frac{1}{\gamma_1}-\int_0^\infty p_1^2 dp_1 
\frac{g_1^2(p_1)}{W_0-\sqrt{\left(\sqrt{m_N^2+p_1^2}+\sqrt{m_\pi^2+p_1^2} 
\right)^2+q_1^2}-\sqrt{m_{\Lambda_c}^2+q_1^2}+{\rm i}\epsilon}. 
\label{eq:A7} 
\end{equation}

\subsection{The $\pi\Lambda_c$ subsystem} 
\label{sec:piY} 

Here, the separable potential 
\begin{equation} 
V_2(p_2,p_2^\prime)=\gamma_2 g_2(p_2) g_2(p_2^\prime), 
\label{eq:B5p} 
\end{equation} 
is used with the form factor 
\begin{equation} 
g_2(p_2)=p_2(1+Ap_2^2)\exp(-p_2^2/\beta_2^2), 
\label{eq:B5} 
\end{equation} 
where the three parameters $\gamma_2$, $\beta_2$ and $A$ were fitted to 
the two pieces of data available, namely, the position and width of the 
$\Sigma_c(2520)$ resonance \cite{pdg}. A family of such fitted parameters 
is given in Table~\ref{tab:piY}. Scanning over $A$ between 0 and 1 gave 
unrealistically small positive values of $<r^2>_{\tilde{g_2}}$ associated 
with the form factor $g_2$, decreasing rapidly with $A$ and becoming negative 
for $A$ exceeding 0.2. Our alternative choice of $r_0$ for a size parameter 
gives values monotonically increasing from 1 to 1.5 upon increasing $A\neq 0$. 
Anticipating $r_0^{(\pi\Lambda_c)}$ to somewhat exceed $r_0^{(\pi N)}=1.36$~fm 
(Table~\ref{tab:piN}), because the pionic $\Lambda_c \to \Sigma_c$ $p$-wave 
excitation energy of 231.5~MeV is smaller than the corresponding excitation 
energy 293~MeV for $N\to\Delta$, we consider the last two rows in 
Table~\ref{tab:piY} as the most physically acceptable fits. For further 
discussion of form-factor sizes, see Ref.~\cite{GAGA11}. 

\begin{table}[hbt] 
\caption{Fitted parameters of the $\pi\Lambda_c$ separable $p$-wave 
interaction defined by Eqs.~(\ref{eq:B5p}) and (\ref{eq:B5}), for chosen 
values of the parameter $A$. Listed also are values of the r.m.s. 
momentum $\sqrt{<p^2>_{g_2}}$ (in fm$^{-1}$), the r.m.s. radius 
$\sqrt{<r^2>_{\tilde{g_2}}}$ and zero $r_0^{(\pi\Lambda_c)}$ (both in fm) 
of the Fourier transform ${\tilde {g_2}}(r)$.} 
\begin{ruledtabular} 
\begin{tabular}{cccccc} 
$A~({\rm fm}^2)$ & $\gamma_2~({\rm fm}^{4})$ & $\beta_2~({\rm fm}^{-1})$ & 
$\sqrt{<p^2>_{g_2}}$ & $\sqrt{<r^2>_{\tilde{g_2}}}$ & $r_0^{(\pi\Lambda_c)}$ 
\\  \hline 
0.0 & $-$0.0044983  & 6.4738   & 9.155 & 0.437 & -- \\ 
0.1 & $-$0.0057655  & 3.6951   & 6.108 & 0.332 & 0.973 \\ 
0.2 & $-$0.0062314  & 3.1432   & 5.258 & 0.070 & 1.103 \\ 
0.3 & $-$0.0063429  & 2.8568   & 4.806 &  --   & 1.194 \\ 
0.4 & $-$0.0062715  & 2.6737   & 4.515 &  --   & 1.263 \\ 
0.5 & $-$0.0061012  & 2.5441   & 4.307 &  --   & 1.318 \\ 
0.7 & $-$0.0056362  & 2.3695   & 4.026 &  --   & 1.401 \\ 
1.0 & $-$0.0048792  & 2.2121   & 3.772 &  --   & 1.487 \\ 
\end{tabular} 
\end{ruledtabular} 
\label{tab:piY} 
\end{table}

\section{The $\Lambda_c N$ subsystem} 
\label{sec:YN} 

There is no experimental data on the $\Lambda_c N$ subsystem that one may rely 
upon to fit a separable potential form. Therefore, and as a guide, we have 
generated local potentials in the $I=\frac{1}{2},{^3S_1}$ channel from the 
recent application of the CCQM to the charmed meson sector \cite{Car12}. 
A brief description of the essential properties required in this model to 
provide interaction output for the $\Lambda_c N$ system follows. 

\subsection{Extension of the CCQM to the charm sector} 
\label{sec:CCQM} 

Baryons are described in the CCQM as clusters of three interacting massive 
constituent quarks, with the light-quark ($u,d$) mass generated by the 
spontaneously broken $SU(2)_{L}\otimes SU(2)_{R}$ chiral symmetry of the QCD 
Lagrangian. Hence, light quarks interact nonperturbatively via Nambu-Goldstone 
boson-exchange potentials 
\begin{equation}
V_{\chi}(\vec{r}_{ij})\, = \, V_{\rm OSE}(\vec{r}_{ij}) \, + \, 
V_{\rm OPE}(\vec{r}_{ij}) \, ,
\label{eq:chi} 
\end{equation} 
given in obvious notation by 
\begin{eqnarray}
V_{\rm OSE}(\vec{r}_{ij}) &=&
    -\dfrac{g^2_{\rm ch}}{{4 \pi}} \,
     \dfrac{\Lambda^2}{\Lambda^2 - m_{\sigma}^2}
     \, m_{\sigma} \, \left[ Y (m_{\sigma} \,
r_{ij})-
     \dfrac{\Lambda}{{m_{\sigma}}} \,
     Y (\Lambda \, r_{ij}) \right] \,,  \\
\label{eq:OSE} 
V_{\rm OPE}(\vec{r}_{ij})&=&
     \dfrac{ g_{\rm ch}^2}{4
\pi}\dfrac{m_{\pi}^2}{12 m_i m_j}
     \dfrac{\Lambda^2}{\Lambda^2 - m_{\pi}^2}
m_{\pi}
     \Biggr\{\left[ Y(m_{\pi} \,r_{ij})
     -\dfrac{\Lambda^3}{m_{\pi}^3} Y(\Lambda
\,r_{ij})\right]
     \vec{\sigma}_i \cdot \vec{\sigma}_j 
\nonumber \\
&&   \qquad\qquad +\left[H (m_{\pi} \,r_{ij})
     -\dfrac{\Lambda^3}{m_{\pi}^3} H(\Lambda
\,r_{ij}) \right] S_{ij}
     \Biggr\}  \vec{\tau}_i \cdot \vec{\tau}_j\, , 
\label{eq:OPE} 
\end{eqnarray}
where $g^2_{\rm ch}/4\pi$ is the chiral coupling constant, $Y(x)$ is 
the Yukawa function, $Y(x)=e^{-x}/x$, and $H(x)=(1+3/x+3/x^2)\,Y(x)$ 
is associated with the quark-quark tensor operator 
$S_{ij} \, = \, 3 \, ({\vec \sigma}_i \cdot {\hat r}_{ij}) 
({\vec \sigma}_j \cdot  {\hat r}_{ij})\, - \, 
{\vec \sigma}_i \cdot {\vec \sigma}_j$. 
The values used for the mass, coupling-constant and cut-off parameters 
are listed in Table 2 of \cite{Car12}. In the case of the heavy charmed quark 
$c$, for which chiral symmetry is explicitly broken, no boson-exchange is 
operative in its interactions with the other quarks. 

Perturbative effects within QCD are accounted for by the one-gluon-exchange 
(OGE) potential 
\begin{equation} 
V_{\rm OGE}({\vec{r}}_{ij})=
        {\frac{\alpha_s}{4}}\,{\vec{\lambda}}_{i}^{\rm c} 
\cdot {\vec{\lambda}}_{j}^{\rm c}
        \left[ {\frac{1}{r_{ij}}}
        - \dfrac{1} {4} \left(
{\frac{1}{{2\,m_{i}^{2}}}}\, +
{\frac{1}{{2\,m_{j}^{2}}}}\,
        + {\frac{2 \vec \sigma_i \cdot \vec
\sigma_j}{3 m_i m_j}} \right)\,\,
          {\frac{{e^{-r_{ij}/r_{0}}}}
{{r_{0}^{2}\,\,r_{ij}}}}
        - \dfrac{3 S_{ij}}{4 m_q^2 r_{ij}^3}
        \right]\,\, ,
\label{eq:OGE} 
\end{equation} 
where $\lambda^{c}$ are the $SU(3)$ color matrices, $r_0$ is 
a flavor-dependent regularization that scales with the reduced mass of the 
interacting pair, and $\alpha_s$ is the QCD scale-dependent coupling constant 
which assumes values of $\alpha_s\sim0.54$ for light-quark pairs and 
$\alpha_s\sim0.43$ for $uc$ and $dc$ pairs \cite{Vij05}. 

Finally, to fully simulate QCD one needs to incorporate confinement. 
While negligible for hadron-hadron interactions, lattice calculations 
suggest that the confinement potential is screened upon increasing the 
interquark distance~\cite{Bal01}, 
\begin{equation}
V_{\rm CON}(\vec{r}_{ij})=[-a_{c}\,(1-e^{-\mu_c\,r_{ij}})]
\vec{\lambda^c}_{i}\cdot \vec{\lambda^c}_{j}\, , 
\label{eq:CON} 
\end{equation} 
with a scale given by $a_c=230$~MeV and a screening mass identified here 
with the pion mass: $\mu_c=m_{\pi}$. 

The CCQM yields a good description of meson~\cite{Vij05} and baryon 
spectra~\cite{Val05b}. Furthermore, by applying resonating-group methods 
it enables one to derive baryon-baryon ($BB$) potentials and, in particular, 
to reproduce the main features of the $NN$ interaction \cite{Val05}. 
Thus, the $B_n B_m\to B_k B_l$ local transition potential 
$V_{B_n B_m (L\, S\, T)\rightarrow B_k B_l (L^{\prime}\, S^{\prime}\, T)}(R)$ 
is derived within a Born-Oppenheimer approximation as 
\begin{equation} 
V_{B_n B_m (L\, S\, T)\rightarrow B_k B_l (L^{\prime}\, S^{\prime}\, T)}(R) = 
\xi_{L \,S \, T}^{L^{\prime}\, S^{\prime}\, T} (R) \, - \, \xi_{L \,S \,
T}^{L^{\prime}\, S^{\prime}\, T} (\infty) \, ,  
\label{eq:BOP} 
\end{equation} 
where
\begin{equation}
\xi_{L\, S\, T}^{L^{\prime}\, S^{\prime}\, T}(R)\, = \, {\frac{{\left
\langle \Psi_{B_k B_l}^{L^{\prime}\, S^{\prime}\, T} ({\vec R}) \mid
\sum_{i<j=1}^{6} V_{q_iq_j}({\vec r}_{ij}) \mid \Psi_{B_n B_m}^{L\, S\, T}
({\vec R})\right\rangle} }{{\sqrt{\left \langle \Psi_{B_k B_l }^{L^{\prime}
\,S^{\prime}\, T}({\vec R})\mid\Psi_{B_k B_l }^{L^{\prime}\, S^{\prime}\, T} 
({\vec R}) \right \rangle} \sqrt{\left \langle \Psi_{B_n B_m }^{L \, S \, T} 
({\vec R})\mid\Psi_{B_n B_m }^{L\, S\, T}({\vec R}) \right \rangle}}}} \, ,
\label{eq:xi} 
\end{equation} 
with the quark coordinates integrated out. The wavefunction 
$\Psi_{B_n B_m}^{L \, S \, T}({\vec R})$ for the two-baryon system is an 
antisymmetrized product of two three-quark clusters, each cluster consisting 
of Gaussian quark wavefunctions. The Gaussian size parameters from Table 2 
in Ref.~\cite{Car12} are $b_n=0.518$~fm for the ($u,d$) light quarks, here 
denoted $n$, and $b_c=0.6$~fm for the $c$ quark. However, whereas the value 
adopted for $b_n$ was deduced long ago by fitting to the $NN$ phase shifts 
and the deuteron binding energy \cite{Val94} (see also the discussion 
in Ref.~\cite{Val96}), the value $b_c=0.6$~fm is not constrained by any 
comparable $BB$ data. It was argued in Ref.~\cite{Car11} that a considerably 
smaller value of $b_c$, in fact $b_c\approx 0.2$~fm, is required to describe 
consistently doubly charmed exotic mesons. Such a value may also be justified 
by recalling that $b_q$ scales with quark-mass as $b_q\sim m_q^{-1/2}$ for 
harmonic-oscillator quark potential. For CCQM quark masses $m_n=313$~MeV and 
$m_c=1752$~MeV, the widely adopted value $b_n=0.518$~fm implies that $b_c=b_n
(m_n/m_c)^{1/2}=0.219$~fm. This strong dependence on the constituent quark 
mass by far overshadows the weak flavor dependence of the harmonic-oscillator 
$1\hbar\omega$ excitation energy, of order hundreds of MeV, in mesons and in 
baryons.  
%

\subsection{The CCQM $I=\frac{1}{2},{^3S_1}$ $\Lambda_c N$ interaction} 
\label{sec:LcN} 

\begin{figure}[htb]
\begin{center}  
\includegraphics[width=0.48\columnwidth]{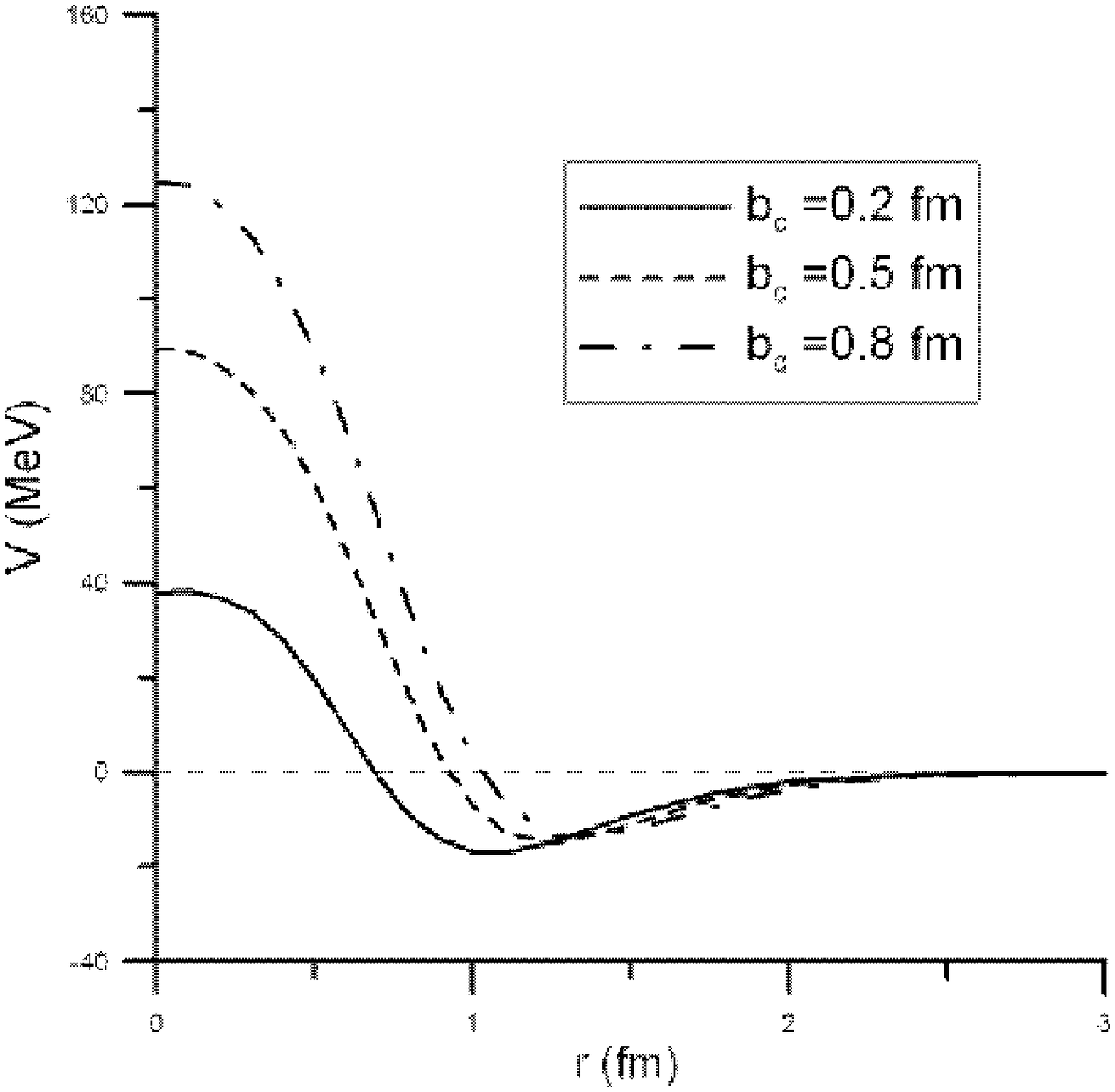}
\includegraphics[width=0.48\columnwidth]{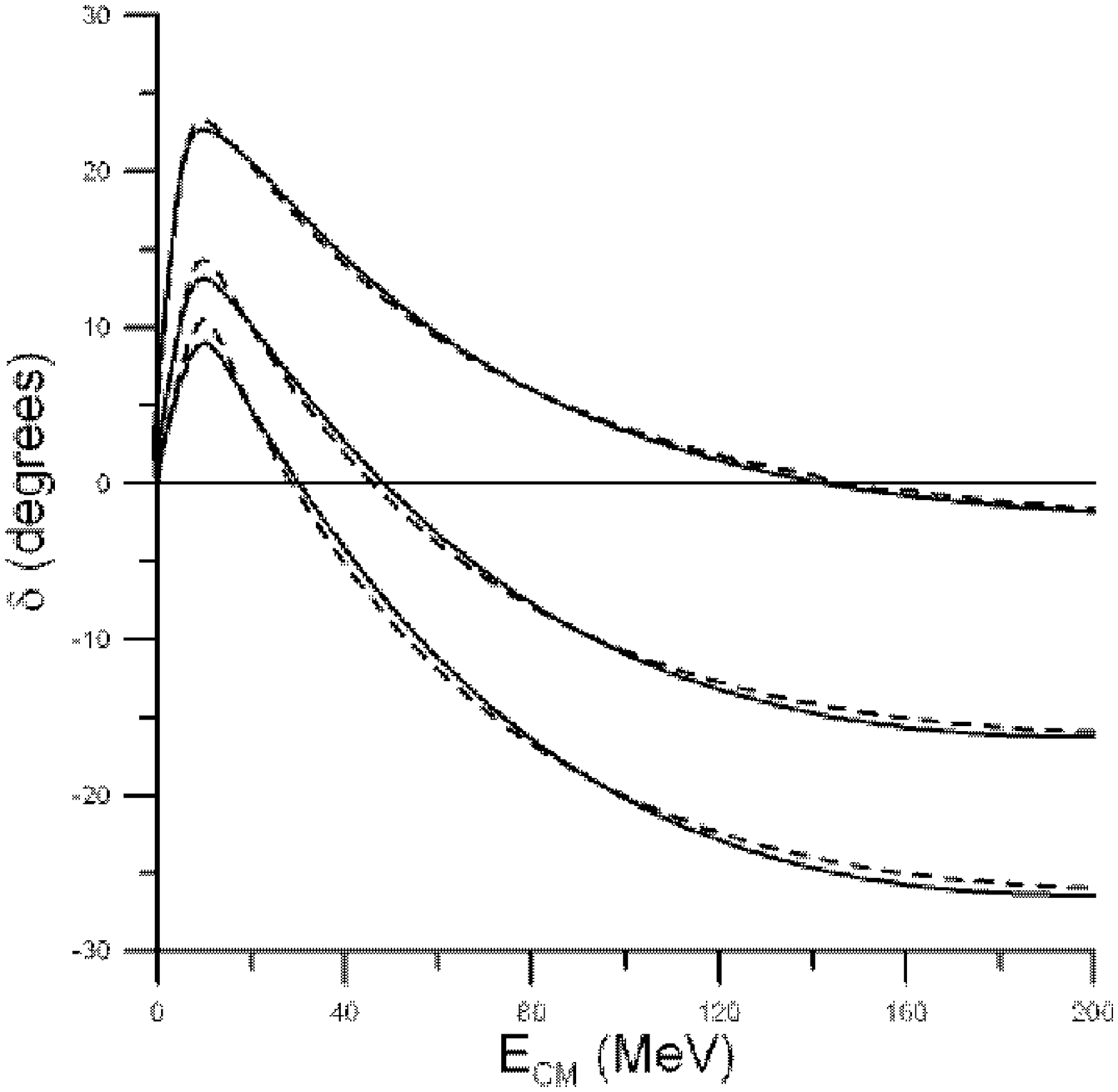} 
\end{center} 
\caption{Left: CCQM $\Lambda_c N$ local potentials in the $^3S_1$ channel for 
several values of the charmed quark harmonic oscillator size parameter $b_c$. 
Right: corresponding $\Lambda_c N$ $^3S_1$ phase shifts, in solid (dashed) 
lines, produced by the CCQM local (separable) potentials. The smaller $b_c$ 
is, the higher is the maximum of the phase shift $\delta$.} 
\label{fig:V3S1} 
\end{figure} 

Adopting the CCQM $s$-wave potentials for the $\Lambda_c N$ interacting pair, 
we show on the l.h.s. of Fig.~\ref{fig:V3S1} three such potentials for the 
$I=\frac{1}{2},{^3S_1}$ channel, using three different values of the charmed 
quark oscillator parameter $b_c$ within the six-quark wave function: 
$b_c=0.2$~fm, $b_c=0.5$~fm and $b_c=0.8$~fm. It is seen that the model with 
$b_c=0.2$ fm has the softest repulsive core and the model with $b_c=0.8$ fm 
has the strongest repulsion at short distance. None of these models is any 
close to generating a $\Lambda_c N$ bound state. The $^3S_1$ phase shifts 
produced by these potentials are shown by the solid lines on the r.h.s. of 
the figure where the change of sign of the phase shift for the model with 
$b_c=0.2$ fm occurs at c.m. energy similar to that of other baryon-baryon 
systems in the CCQM such as $NN$ or $\Lambda N$, while for the models with 
$b_c=0.5$ and $b_c=0.8$ fm the change of sign occurs at very low c.m. energy 
which suggests that these latter two models have excessive repulsion. We also 
show by the dashed lines the phase shifts obtained using the rank-2 separable 
potential with both attraction and repulsion 
\begin{equation} 
V_3(p_3,p_3^\prime)=- g_3^a(p_3) g_3^a(p_3^\prime) 
                    + g_3^r(p_3) g_3^r(p_3^\prime), 
\label{eq:B6} 
\end{equation} 
so that the corresponding two-body $t$-matrix is given by 
\begin{equation} 
t_3(p_3,p_3^\prime;\omega_0)=-\sum_{\alpha=a,r}\sum_{\beta=a,r} 
g_3^\alpha(p_3)\tau_3^{\alpha\beta}(\omega_0)g_3^\beta(p_3^\prime)~, 
\label{eq6} 
\end{equation} 
where 
\begin{equation} 
\tau_3^{ar}(\omega_0)=\tau_3^{ra}(\omega_0)=\frac{G_3^{ar}(\omega_0)}
{[1+G_3^{aa}(\omega_0)][1-G_3^{rr}(\omega_0)]+[G_3^{ar}(\omega_0)]^2}\;, 
\label{eq7} 
\end{equation} 
\begin{equation} 
\tau_3^{aa}(\omega_0)=\frac{1-G_3^{rr}(\omega_0)}{[1+G_3^{aa}(\omega_0)] 
[1-G_3^{rr}(\omega_0)]+[G_3^{ar}(\omega_0)]^2}\;, 
\label{eq8} 
\end{equation} 
\begin{equation} 
\tau_3^{rr}(\omega_0)=-\frac{1+G_3^{aa}(\omega_0)}{[1+G_3^{aa}(\omega_0)] 
[1-G_3^{rr}(\omega_0)]+[G_3^{ar}(\omega_0)]^2}\;, 
\label{eq9} 
\end{equation} 
with $G_3^{\alpha\beta}(\omega_0)$ given by 
\begin{equation} 
G_3^{\alpha\beta}(\omega_0)=\int_0^\infty p_3^2dp_3\frac{g_3^\alpha(p_3) 
g_3^\beta(p_3)}{\omega_0-\sqrt{p_3^2+m_N^2}-\sqrt{p_3^3+m_{\Lambda_c}^2}+
i\epsilon}\;. 
\label{eq10} 
\end{equation} 
The form factors $g_3^\beta(p_3)$ 
are chosen to be of the Yamaguchi form 
\begin{equation} 
g_3^\beta(p_3)=\frac{\sqrt{\gamma_\beta}}{p_3^2+\alpha_\beta^2}~~~~
(\beta=a,r)~,
\label{eq11} 
\end{equation} 
and the parameters of these models are given in Table \ref{tab:YN} together 
with values of the associated scattering lengths and effective ranges. The 
relatively small size of the scattering lengths $a_{\Lambda_c N}$ clearly 
indicates that the $^3S_1$ $\Lambda_c N$ system is far from binding on its 
own. 

\begin{table}[htb] 
\caption{Parameters of the $^3S_1$ $\Lambda_c N$ separable potential models 
Eqs.~(\ref{eq:B6}), (\ref{eq11}).} 
\begin{ruledtabular} 
\begin{tabular}{ccccccc} 
$b_c$ (fm) & $\gamma_3^a$ (fm$^2$) & $\alpha_3^a$ (fm$^{-1}$) 
& $\gamma_3^r$ (fm$^2$) & $\alpha_3^r$ (fm$^{-1}$) & $a$ (fm) & $r$ (fm) \\ 
\hline 
0.2 & 1.8915 & 1.7672 & 2.6210 & 2.1523 & $-$1.33 & 3.3 \\ 
0.5 & 2.1804 & 1.7030 & 4.5651 & 2.1973 & $-$0.79 & 5.6 \\ 
0.8 & 2.3435 & 1.6286 & 5.7197 & 2.1338 & $-$0.63 & 8.2 \\ 
\end{tabular} 
\end{ruledtabular} 
\label{tab:YN} 
\end{table}

\section{Results and discussion} 
\label{sec:res} 

Solutions of the Faddeev equations corresponding to bound states and resonance 
poles in the $(I,J^P)=(\frac{3}{2},2^+)$ channel of the $\pi\Lambda_c N$ 
three-body system were found applying search procedures described in 
Refs.~\cite{GAGA08,GAGA10,GAGA13}. A single bound state or resonance was 
established for any combination of each one of the $\pi\Lambda_c$ interaction 
models specified in Table~\ref{tab:piY} and each one of the $\Lambda_c N$ 
interaction models specified in Table~\ref{tab:YN}, as well as for the case 
when there is no $\Lambda_c N$ interaction. The resulting bound-state and 
resonance energies are given in Table~\ref{tab:res} with respect to the 
$\pi\Lambda_c N$ threshold mass $E_{\rm th}\approx 3363$~MeV.  

\begin{table}[ht] 
\caption{Energy eigenvalue of the $(I,J^P)=(\frac{3}{2},2^+)$ $\pi\Lambda_c N$ 
state (in MeV with respect to the $\pi\Lambda_c N$ threshold) for the eight 
models of the $\pi\Lambda_c$ interaction characterized by the parameter $A$ 
and the three models of the $\Lambda_c N$ interaction $b_c=0.2$, $b_c=0.5$, 
and $b_c=0.8$~fm, as well as for $V_{\Lambda_c N}=0$.}
\begin{ruledtabular} 
\begin{tabular}{ccccccc} 
& $A~({\rm fm}^2)$ & $b_c=0.2$&$b_c=0.5$& $b_c=0.8$& $V_{\Lambda_c N}=0$ & \\ 
\hline  
& 0.0 & $-$2.3       & 36.7$-$i0.45 & 63.1$-$i1.91 & $-$0.7 & \\  
& 0.1 & $-$14.6      & 26.3$-$i0.18 & 54.8$-$i1.16 & $-$12.5 & \\ 
& 0.2 & $-$12.7      & 27.2$-$i0.21 & 55.2$-$i1.22 & $-$10.1 & \\ 
& 0.3 & $-$8.4       & 30.0$-$i0.28 & 57.0$-$i1.53 & $-$5.3 & \\ 
& 0.4 & $-$4.0       & 33.1$-$i0.38 & 59.1$-$i1.67 & $-$0.5 & \\  
& 0.5 & 0.1$-$i0.00  & 35.9$-$i0.51 & 61.2$-$i1.96 & 3.9$-$i0.00 & \\ 
& 0.7 & 7.0$-$i0.01  & 40.8$-$i0.80 & 64.4$-$i2.34 & 11.3$-$i0.01 & \\ 
& 1.0 & 14.7$-$i0.07 & 46.2$-$i1.10 & 68.2$-$i3.05 & 19.5$-$i0.07 & \\  
\end{tabular} 
\end{ruledtabular} 
\label{tab:res} 
\end{table} 

Bound-state solutions appear in several of the $b_c=0.2$~fm models and also 
when the $\Lambda_c N$ interaction is switched off, whereas the models 
$b_c=0.5$ and $b_c=0.8$~fm give only resonance solutions. However, when the 
resonance lies below the $\Sigma_c N$ threshold (Re~$E<$ 27 MeV) the resonance 
states are quite narrow with widths less than 0.4 MeV. This does not apply 
to the models with $b_c=0.5$ and $b_c=0.8$~fm in which the resonance lies 
above the $\Sigma_c(2455)N$ threshold and its width is therefore larger than 
indicated by the tabulated widths. 

As concluded in Sect.~\ref{sec:piY}, only values of $A > 0.5$~fm$^2$ are 
acceptable in considering the $\Lambda_c(2286)+\pi\to\Sigma_c(2520)$ $p$-wave 
form factor relative to the $N(939)+\pi\to\Delta(1232)$ $p$-wave form factor. 
Combined with the more plausible $\Lambda_c(2286)N$ interaction model 
defined by choosing $b_c=0.2$~fm, or even neglecting for simplicity this 
$\Lambda_c N$ interaction, we conclude from Table~\ref{tab:res} that 
the $(I,J^P)=(\frac{3}{2},2^+)$ state of the $\pi\Lambda_c N$ system 
resonates at energy up to about 20 MeV above threshold, that is at 
$3363\lesssim\sqrt{s}\lesssim 3383$~MeV. This resonance is expected to be 
quite narrow, with width less than 0.2 MeV dominated by its elastic width. 
Pion absorption can occur only by violating charge independence, 
($I_{\pi\Lambda_c N}=\frac{3}{2}$)$\to$($I_{\Lambda_c N}=\frac{1}{2}$), 
and the lowest $\Sigma_c N$ channel is closed at the energy range expected 
for the resonance.

\section{Conclusion} 
\label{sec:concl} 

It was demonstrated in this work by solving Faddeev equations that the 
$(I,J^P)=(\frac{3}{2},2^+)$ state of the $\pi\Lambda_c N$ system is 
a strong candidate for a pion-assisted charmed dibaryon. Although the CCQM 
$\Lambda_c N$ interaction is not sufficiently strong to bind a $\Lambda_c N$ 
$^3S_1$ pair (and this holds even more so for a $^1S_0$ pair in the CCQM), 
the $p$-wave pion attractive interactions induced by the $J^P={\frac{3}{2}}^+$ 
$\Delta(1232)$ and $\Sigma_c(2520)$ resonances manage to bind the 
$\pi\Lambda_c N$ three-body system or more likely to make it resonate. 
The prediction of this dibaryon candidate is robust in the sense that 
its existence depends little on the $\Lambda_c N$ spin-triplet $s$-wave 
interaction, even if the precise energy of the resonance is not pinned 
down between threshold at $\approx$3363~MeV and several tens of MeV above 
threshold according to the variation offered in Table~\ref{tab:res}. 
This resonance is likely to be the {\it lowest lying} charmed dibaryon, 
considerably below the mass $\approx$3500~MeV predicted recently for 
a $DNN$ bound state with quantum numbers $I=\frac{1}{2},J^P=0^-$ that 
may be viewed also as a $\Lambda_c(2595)N$ bound state \cite{DNN12}. 
These two charmed-dibaryon predictions, with $(I,J^P)=(\frac{3}{2},2^+)$ 
(ours) versus $(I,J^P)=(\frac{1}{2},0^-)$ \cite{DNN12}, bear structural 
resemblance to the strange-dibaryon predictions of $\pi\Lambda N(I=\frac{3}{2}
,J^P=2^+)$ \cite{GAGA08,GAGA10,GAGA13} versus $K^-pp(I=\frac{1}{2},J^P=0^-)$ 
that may also be viewed as a $\Lambda(1405)N$ quasibound state \cite{Gal13}. 
None of these dibaryon candidates has been confirmed by experiment. 

Denoting the $(I,J^P)=(\frac{3}{2},2^+)$ $\pi\Lambda_c N$ dibaryon candidate 
by ${\cal Y}_c$, in analogy to the $(I,J^P)=(\frac{3}{2},2^+)$ $\pi\Lambda N$ 
dibaryon candidate $\cal Y$ \cite{Gal13}, the following production reactions 
of ${\cal Y}_c$ are feasible with proton and pion beams in the high-momentum 
hadron beam line extension approved at J-PARC: 
\begin{eqnarray} 
   p~ + ~p & ~\rightarrow ~ & {\cal Y}_c^{+++} ~+~D^- \nonumber  \\  
           &                & ~\hookrightarrow ~ \Sigma_c^{++}(2455)~+~ p \; ,
\label{eq:pptoYc+++} 
\end{eqnarray} 
\begin{eqnarray} 
\pi^+ ~+~ d & ~\rightarrow ~ & {\cal Y}_c^{+++} ~+~D^-  \nonumber  \\ 
 &  &  ~\hookrightarrow ~ \Sigma_c^{++}(2455)~+~ p \; , 
\label{eq:pi+dtoYc+++}  
\end{eqnarray} 
\begin{eqnarray} 
\pi^- ~+~ d & ~\rightarrow ~ & {\cal Y}_c^+ ~+~D^-  \nonumber  \\ 
 &  &  ~\hookrightarrow ~ \Sigma_c^{+/0}(2455)~+~ n/p \; .  
\label{eq:pi-dtoYc+}  
\end{eqnarray} 
The ${\cal Y}_c$ dibaryon resonance may be looked for both within inclusive 
missing-mass measurements by focusing on the outgoing $D^-$ charmed meson, 
and in exclusive invariant-mass measurements focusing on the outgoing 
$\Sigma_c(2455)N$ decay pair provided that ${\cal Y}_c$ is located above 
the $\Sigma_c(2455)N$ threshold.

\acknowledgments 

The research of A.G. is partially supported by the HadronPhysics3 networks 
SPHERE and LEANNIS of the European FP7 initiative, the research of H.G. is 
supported in part by COFAA-IPN (M\'exico), and the research of A.V. and 
T.F.C. is supported by the Spanish Ministerio de Educaci\'on y Ciencia 
and EU FEDER under Contract No. FPA2010-21750, and by the Spanish 
Consolider-Ingenio 2010 Program CPAN (CSD2007-00042).

\end{document}